\documentclass[preprint,5p,times,twocolumn]{elsarticle}
\usepackage{amsmath}
\usepackage{graphicx}
\usepackage{multirow}
\usepackage{amssymb}
\usepackage{footnote}
\usepackage{lineno}
\usepackage{epsfig}
\usepackage{url}

\newcommand{\natte}{\mbox{$^{nat}$Te} }
\newcommand{\zn}{\mbox{0$\nu\beta\beta$}}
\makeatletter
\def\ps@pprintTitle{%
 \let\@oddhead\@empty
 \let\@evenhead\@empty
 \def\@oddfoot{}%
 \let\@evenfoot\@oddfoot}
\makeatother
\begin{document}

\title{Cosmogenic activation of a natural tellurium target}

\author[1]{V. Lozza\corref{cor1}}
\ead{valentina.lozza@tu-dresden.de}
\author[1]{J. Petzoldt\fnref{fn2}}
\cortext[cor1]{Corresponding author: V. Lozza}
\fntext[fn2]{Present address: OncoRay, National Center for Radiation Research in Oncology, 01307 Dresden, Germany}
\address[1]{Institut f\"{u}r Kern und Teilchenphysik, Technische Universit\"{a}t Dresden, Zellescher Weg 19, 01069 Dresden, Germany}

\begin{abstract}
$^{130}$Te is one of the candidates for the search for neutrinoless double beta decay. It is currently planned to be used in two experiments: CUORE and SNO+. In the CUORE experiment TeO$_{2}$ crystals cooled at cryogenic temperatures will be used. In the SNO+ experiment $^{nat}$Te will be deployed up to 0.3\% loading in the liquid scintillator volume. A possible background for the signal searched for, are the high Q-value, long-lived isotopes, produced by cosmogenic neutron and proton spallation reaction on the target material. A total of 18 isotopes with Q-value larger than 2 MeV and T$_{1/2}>$20 days have been identified as potential backgrounds. In addition low Q-value, high rate isotopes can be problematic due to pile-up effects, specially in liquid scintillator based detectors. Production rates have been calculated using the ACTIVIA program, the TENDL library, and the cosmogenic neutron and proton flux parametrization at sea level from Armstrong and Gehrels for both long and short lived isotopes. The obtained values for the cross sections are compared with the existing experimental data and calculations. Good agreement has been generally found. The results have been applied to the SNO+ experiment for one year of exposure at sea level. Two possible cases have been considered: a two years of cooling down period deep underground, or a first purification on surface and 6 months of cooling down deep underground. Deep underground activation at the SNOLAB location has been considered.
\end{abstract}



\maketitle
\section{Introduction}
The neutrino was introduced for the first time by Pauli in 1930 as a solution for the beta-decay spectrum. Initially considered as a massless particle, during the period from 1990 to 2006 especially the SNO \cite{SNO09} and the SuperK \cite{SuperK98} oscillation experiments proved that neutrinos change their flavor when traveling from the source to the detector. This property was already known in the quark sector and happens only if the particles involved do have mass. The existence of massive neutrinos has been introduced within extensions of the Standard Model. Two types of neutrino masses are allowed, the Dirac and the Majorana ones. In the Majorana case neutrinos and their anti-particles are the same. \\
The investigation of the neutrino's nature is one of the most active fields in modern neutrino physics. The search of the Majorana nature of neutrinos is done in the Double-Beta-Decay (DBD) experiments \cite{Rod12}. The DBD is a nuclear process where the Z number changes by 2 units while the atomic mass, A, does not change. It happens only if the single beta decay transition is forbidden or strongly suppressed. If neutrinos are Majorana particles, the decay can proceed without the emission of the 2 neutrinos. The result is the presence of a peak at the Q-value of the reaction in the sum energy spectrum of the two electrons.\\
There are about 35 candidates nuclei for the DBD searches. The quantity extracted from the number of events in the peak is the half-life of the decay. The expected value for the neutrinoless double beta decay half life is larger than 10$^{25}$ years. To search for such a rare event a large mass (volume), a very low background environment and a good knowledge of the backgrounds in the detector are necessary.\\
A possible source of background is the activation of the candidate material through spallation reaction induced by cosmogenic neutrons and protons during handling on surface. A study of the cosmogenic-induced isotopes has been done for several double beta decay experiments, like GERDA \cite{Bar06} and CUORE \cite{CUORE08}, in order to define the maximum allowed exposure on surface, the necessary shielding during transportation, the necessary purification factors, and the cooling down time underground (UG) to reduce the cosmogenic background to a negligible level.\\
In this article the studies are focused on a natural tellurium target. $^{130}$Te is the candidate nucleus chosen by the CUORE and the SNO+ collaborations for the search for the neutrinoless double beta decay. $^{130}$Te has a high natural isotopic abundance of 34.08\% and a relative high Q-value of 2.53 MeV. The CUORE collaboration will use tellurium in form of crystals (TeO$_{2}$) cooled at cryogenic temperatures. The energy resolution of the crystals is nearly 5--7\,keV (FWHM) at 2.53\,MeV \cite{Vig13}. The cosmogenic-induced isotopes identified as possible background by the collaboration are $^{60}$Co, $^{124}$Sb and $^{110m}$Ag. The calculated exposure time at sea level for the Te crystals in \cite{CUORE08} was 4 months, followed by 2 years of storage underground.\\
The SNO+ experiment will use a tellurium loaded liquid scintillator to search for neutrinoless double beta decay. The tellurium acid (Te(OH)$_{6}$) is dissolved in the liquid scintillator using water and a surfactant \cite{Bil13}. The SNO+ energy resolution is expected to be worse than the CUORE one. An energy resolution of 220--270\,keV (FWHM) at 2.53\,MeV can be estimated using the 200--300\,pe/MeV as in \cite{Bil13}. The actual resolution of the SNO+ experiment will depend on the final composition of the scintillator cocktail.
Due to the worse energy resolution than bolometer experiments a larger number of cosmogenic-induced background nuclides can possibly fall in the energy region where the 0$\nu\beta\beta$ is expected.\\
The identification of all potential backgrounds contributing in the 0$\nu\beta\beta$ energy region is extremely important to claim a discovery.\\ 
The aim of this paper is to find potential cosmogenic-induced background candidates for Te-based experiments. For each isotope the production rate is calculated. The actual danger level of the identified candidates should be evaluated by the different experiments depending on the energy resolution, the target background level in the region of interest and the specific exposure and cooling down time.

\section{Isotopes}

The list of potential background candidates, for the double beta decay search in Te-based experiment, has been obtained searching into the table of isotopes (ToI)\cite{fir14} for candidates that match the following criteria:
\begin{itemize}
\item T$_{1/2}>$ 20 days. Shorter lived isotopes are generally considered to be removed by a cooling down time of few months. A cooling down time of about 6 months will reduce the initial activity of at least a factor 560 for a 20 days lived isotope. Shorter lived isotopes are considered when fed by a long-lived parent. 
\item Q-value larger than 2 MeV. For internal backgrounds both bolometer and scintillator experiments measure the total energy released in the decay (beta and gammas). The Q-value of the reaction is then an important criterion to select potential background isotopes. For an energy resolution of about 250 keV (FHWM) isotopes with Q-value larger than 2.3 MeV can potentially fall in the 0$\nu\beta\beta$ region. To account for a possible worse resolution a minimum of 2\,MeV has been chosen.
\item Mass number, A, smaller than $^{131}$I, since the isotopes are supposed to be produced by spallation reactions on Te. Isotopes with higher mass number can be produced only if contaminations are present in the target material.
\end{itemize}
In some cases, isotopes are both directly produced in the reaction, and fed by the decay of a long-lived parent also produced in the target material. If the half-life of the daughter is much shorter than the parent one, secular equilibrium is assumed. Otherwise, both contributions are taken into account. This is for instance the case for $^{88}$Y that can be directly produced in spallation reactions, but also fed by the decay of $^{88}$Zr.
The identified isotopes are summarized in table \ref{tab::isotopes}. The T$_{1/2}$, Q-value and the decay mode are shown. The decay branching ratio (BR) is listed in brackets when different from 100\%. 

\begin{table*}[ht]\centering
\begin{minipage}{14cm}
\begin{tabular}{cccc}
\hline 
Isotope 	&	T$_{1/2}$ \cite{nudat2}	&	Q-value \cite{nudat2} & Decay mode (BR)\\ 
	&	[d]	&	[MeV]  & (\%)\\ \hline
$^{22}$Na 	&	950.6	&	2.84	 &	EC, $\beta^{+}$ \\ 
$^{26}$Al 	&	2.62E+8	&	4.00	  &$\beta^{+}$\\ 
$^{42}$K (direct and daughter of $^{42}$Ar)	&	0.51 (1.20E+4)	&	3.53	&$\beta^{-}$\\ 
$^{44}$Sc (direct and daughter of $^{44}$Ti) 	&	0.17 (2.16E+4)	&	3.65	& EC, $\beta^{+}$\\ 
$^{46}$Sc 	&	83.79	&	2.37 	&	$\beta^{-}$\\
$^{56}$Co 	&	77.2	&	4.57	&	EC, $\beta^{+}$\\  
$^{58}$Co 	&	70.9	&	2.31	 & EC, $\beta^{+}$ \\  
$^{60}$Co (direct and daughter of $^{60}$Fe)	&	1925.27 (5.48E+8)	&	2.82	&	$\beta^{-}$\\  
$^{68}$Ga (direct and daughter of $^{68}$Ge) 	&	4.70E-2 (271)	&	2.92	&	EC, $\beta^{+}$\\  
$^{82}$Rb (daughter of $^{82}$Sr) 	&	8.75E-4 (25.35)	&	4.40 &	EC, $\beta^{+}$ \\  
$^{84}$Rb  &	32.8	&	2.69  &	 EC, $\beta^{+}$ (96.1)\\  
$^{88}$Y (direct and daughter of $^{88}$Zr) 	&	106.63 (83.4)	&	3.62	&	EC, $\beta^{+}$ \\  
$^{90}$Y (direct and daughter of $^{90}$Sr) 	&	2.67 (1.05E+4)	&	2.28	&	$\beta^{-}$\\  
$^{102}$Rh (direct and daughter of $^{102m}$Rh)	\footnote{0.23\% from $^{102m}$Rh IT decay.} &	207.3	&	2.32 &	 EC, $\beta^{+}$ (78)\\  
$^{102m}$Rh &	1366.77	&	2.46	 &	EC (99.77)\\  
$^{106}$Rh (daughter of $^{106}$Ru)	&	3.47E-4 (371.8)	&	3.54	&	$\beta^{-}$ \\  
$^{110m}$Ag &	249.83	&	3.01 & $\beta^{-}$ (98.67)\\  
$^{110}$Ag (daughter of $^{110m}$Ag)\footnote{1.33\% from$^{110m}$Ag IT decay} 	&	2.85E-4	&	2.89	&	$\beta^{-}$(99.70)\\  
$^{124}$Sb 	&	60.2	&	2.90	& $\beta^{-}$ \\  
$^{126m}$Sb (direct and daughter of $^{126}$Sn) 	&	0.01 (8.40E+7)	&	3.69	 &$\beta^{-}$ (86)\\  
$^{126}$Sb (direct and daughter of $^{126m}$Sb)\footnote{14\% from $^{126m}$Sb IT decay.}	&	12.35 (0.01)	&	3.67	&$\beta^{-}$\\  
\hline
\end{tabular}
\end{minipage}
\caption{Possible background nuclides induced by cosmogenic neutrons and protons on a natural Te target. The isotopes have been searched in the \textit{ToI} based on the criteria described in the text. T$_{1/2}$, Q-value and decay modes are shown. When the branching ratio (BR) is different from 100\%, the value is specified in brackets.}
\label{tab::isotopes}
\end{table*}

\subsection{Other possible background isotopes}

Even if having a transition energy below the DBD one, long-lived low Q-value nuclides can be a background for the search for neutrinoless double beta decay if their decay rate is high enough to produce pile-up background. A high production rate is expected for those isotopes that are close to tellurium (Sn, Sb, In, Te and I). Isotopes have been selected based on the table of isotopes \cite{fir14} and the criteria that they are close to tellurium.\\
Isotopes with T$_{1/2}$ shorter than 20 days and high Q-value can affect the measurement if long cooling down time underground are not possible or if their production rates are too high. In the latter case the isotopes close to tellurium may be the one with higher production rates. A summary of the low Q-value isotopes and the short lived ones (minimum half-life of 5 days) based on the hypothesis done, is shown in table \ref{tab::lowQ}.

\begin{table*}[ht]\centering
\scalebox{0.99}{
\begin{tabular}{cccc}
   \hline
Isotope	&	T$_{1/2}$ \cite{nudat2} & Q-value \cite{nudat2} &  Decay mode (BR) \\
& [d] & [MeV] & (\%) \\\hline
$^{121m}$Sn  & 16034 & 0.40 ($\beta^{-}$), 0.006 (IT) & $\beta^{-}$(22.4), IT(77.6)\\ 
$^{123}$Sn  & 129.2  & 1.40 & $\beta^{-}$\\ 
$^{125}$Sn & 9.64  & 2.36 & $\beta^{-}$\\ 
$^{118}$Sb	(fed by $^{118}$Te) & 0.0025 & 3.66 & $\beta^{+}$, EC \\ 
$^{120m}$Sb & 5.76& 2.68 & EC \\ 
$^{125}$Sb & 1008& 0.77 & $\beta^{-}$ \\ 
$^{127}$Sb & 3.85& 1.58 & $\beta^{-}$  \\ 
$^{129}$Sb & 0.18 & 2.38 & $\beta^{-}$  \\ 
$^{118}$Te (feed $^{118}$Sb) & 6& 0.28 & EC \\ 
$^{121}$Te & 19.17 & 1.05 & EC\\ 
$^{121m}$Te & 164.2& 1.35 (EC), 0.29 (IT) & EC, $\beta^{+}$ (11.4), IT (88.6) \\ 
$^{123m}$Te & 119.2 & 0.25 & IT \\  
$^{125m}$Te & 57.4 & 0.14 & IT \\  
$^{127m}$Te (fed by $^{127}$Sb) & 106.1& 0.79 ($\beta^{-}$), 0.09 (IT)  & $\beta^{-}$ (2.4), IT (97.6)  \\ 
$^{129m}$Te  & 33.6 & 1.60 ($\beta^{-}$),  0.11 (IT) & $\beta^{-}$(37), IT(63)\\
$^{125}$I & 59.4 & 0.19 & EC \\  
$^{126}$I & 12.93& 2.16 (EC)/1.26 ($\beta^{-}$) & EC (52.7), $\beta^{-}$ (47.3)  \\ 
$^{129}$I & 5.73E+9& 0.19 & $\beta^{-}$  \\ 
$^{48}$V & 15.97 & 4.01 & $\beta^{+}$, EC\\
$^{106m}$Ag & 8.28 & 3.05 & EC\\ 
\hline
\end{tabular}}
\caption{Short lived isotopes (minimum half-life of 5 days) and low Q-value isotopes that can be produced by cosmogenic activation of a natural Te sample. The low Q-value isotopes have been selected because being close in mass to tellurium can potentially have a high production rate.}
\label{tab::lowQ}
\end{table*}

\section{Cosmogenic flux parametrization at sea level}
\label{sec::flux}

\begin{figure}[h]\centering
\includegraphics[scale=0.48]{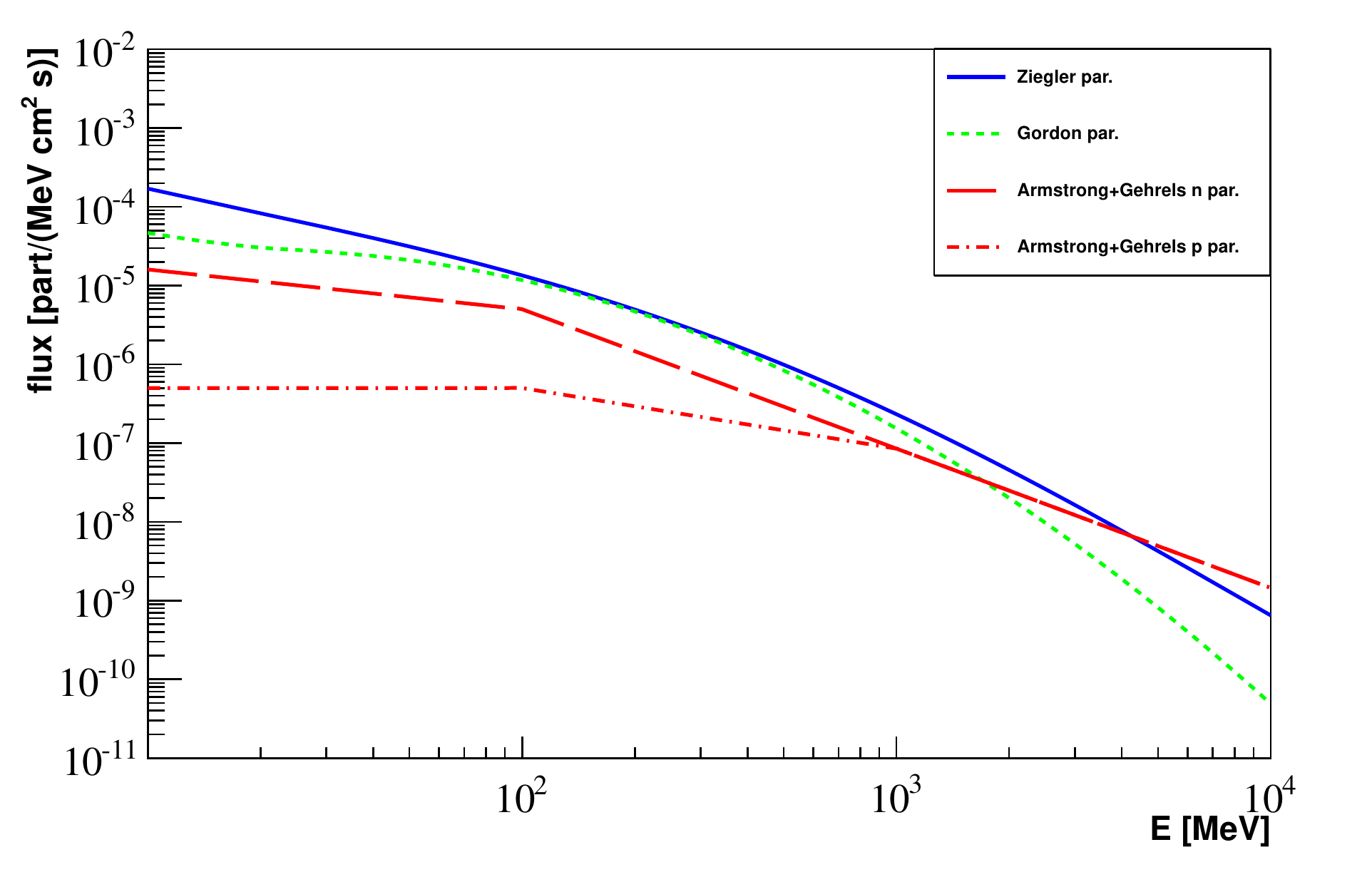}
\caption{Comparison among the different flux parametrizations at sea level: neutron flux parametrization from Gordon \cite{gor04} (green, dashed), neutron (red, long dashed) and proton (red, dashed dotted) flux from Armstrong and Gehrels \cite{arm73}\cite{geh85} and the neutron flux from  Ziegler \cite{zie98} (blue, solid). The Armstrong flux parametrization (neutron+proton) above 3\,GeV is harder than the Ziegler and Gordon neutron ones.}\label{fig::flux}
\end{figure}

At sea level the nucleon component of cosmic rays is mainly neutrons (95\%), protons (3\%) and pions (2\%). Protons, despite the small flux, are harder than neutrons. They are stopped more efficiently by shielding around the target. \\
Different flux parametrizations exist for the neutron and the proton fluxes at sea level. The most common are the one from Ziegler \cite{zie98}, Gordon \cite{gor04} and Armstrong and Gehrels \cite{arm73}\cite{geh85}. A comparison for the different flux parametrizations at sea level is shown in figure \ref{fig::flux} in the energy range from 10\,MeV to 10\,GeV. The parametrizations are generally referred to New York. Correction factors (to be applied to the flux integral) due to location and altitude can be quite important: 0.7 for Beijing, 0.81 for Rome, 1.02 for Greater Sudbury (all referred at 0\,m altitude) and up to 500 for flight altitudes (39000 feet) \cite{nloc}.\\
The flux parametrization at sea-level from Gordon \cite{gor04} is valid only for neutrons in the energy range from 0.4\,MeV to 10\,GeV. The Ziegler parametrization \cite{zie98} is valid for neutrons in the energy range from 10\,MeV to 10\,GeV. Nowadays, there is not a unique parametrization for the cosmic flux at sea-level.  
In the following calculation, the parametrization from Armstrong and Gehrels at sea level is used. This flux parametrization is chosen because it accounts separately for neutrons and protons and it is used in the wide energy range from 10\,MeV to 100\,GeV. \\
Calculation of production rates generally use the Armstrong (modified version of COSMO \cite{cosmo} and ACTIVIA \cite{activia} code) or Ziegler (IDEA evaluation, see section \ref{sec::exc_funcion}) parametrization. Recently the parametrization from Gordon has been also used \cite{Ceb10}.\\ 
The effect of using a different flux parametrization can be quite important. From figure \ref{fig::flux}, it can be seen that the flux parametrization from Gordon and Ziegler dominates in the energy range from 10\,MeV to few hundreds of MeV. The parametrization from Armstrong (neutron+proton) dominates above 3\,GeV. Nuclides that are close in atomic number to tellurium, such as $^{124}$Sb, have generally a high production rate at low energies (\textit{E}$<$100\,MeV). Nuclides that are not close to tellurium, like $^{60}$Co and $^{110m}$Ag, on the contrary, have generally a high production rate at high energies (\textit{E}$>$3\,GeV). In table \ref{tab::percent_change} column 3 and 4  it is shown the change, in per cent, on the expected number of events (t$_{measure}$=1 year) due to the use of the flux parametrization from Ziegler and Gordon respectively. The reference used is the parametrization from Armstrong. In all cases one year of exposure at the sea-level cosmic flux was assumed. The results obtained with either the Ziegler or Gordon parametrizations are higher up to a factor 3 for the isotopes close in atomic number to tellurium. Results are smaller, instead, for isotopes not close to tellurium. It can also be noted from figure \ref{fig::flux}, that the parametrization from Ziegler is closer to the one from Armstrong in the high energy range. As a consequence the results in column 3 of table \ref{tab::percent_change} are less divergent from the one obtained with the parametrization from Armstrong for those isotopes.  The same general conclusion was drawn in \cite{Ceb10} and \cite{IDEA}.

\begin{table*}[ht]\centering
\begin{tabular}{cccc}
  \hline
Isotope 	&	\% Change cross section	&	\% Change flux Ziegler	& \% Change flux Gordon\\ 
	&	\%	&	\%	 & \%\\ \hline
$^{22}$Na &	28.8	&	-52.6	&	-80.9	\\  	
$^{26}$Al &	26.4	&	-52.4	&	-80.7	\\  
$^{42}$K (daughter of $^{42}$Ar)	&	-14.6	&	-35.2	&	-70.8	\\  
$^{44}$Sc (daughter of $^{44}$Ti) &	13.5	&	-52.8	&	-80.8	\\  
$^{46}$Sc &	-22.3	&	-47.4	&	-78.7	\\  
$^{56}$Co &	46.8	&	-51.9	&	-80.2	\\  
$^{58}$Co &	60.2	&	-36.4	&	-69.1	\\  
$^{60}$Co (+ $^{60}$Fe)	&	34.4	&	-34.5	&	-68.9	\\  
$^{68}$Ga (daughter of $^{68}$Ge) &	42.2	&	-35.8	&	-68.4	\\  
$^{82}$Rb (daughter of $^{82}$Sr) &	56.5	&	-31.4	&	-65.4	\\  
$^{84}$Rb &	41.5	&	-31.4	&	-66.4	\\  
$^{88}$Y (+ $^{88}$Zr) &	11.9	&	-28.5	&	-62.9	\\  
$^{90}$Y (daughter of $^{90}$Sr) 	&	-41.2	&	-32.6	&	-68.5	\\  
$^{102}$Rh 	&	17.9	&	12.4	&	-27.4	\\  
$^{102m}$Rh &	17.9	&	12.4	&	-27.4	\\  
$^{106}$Rh (daughter of $^{106}$Ru)	&	32.8	&	24.1	&	-15.5	\\  
$^{110m}$Ag 	&	-2.3(\textit{T})/-1.8	&	50.5	&	14.4	\\  
$^{124}$Sb 	&	-2.37(\textit{T})/-0.04	&	183.6	&	143.0	\\  
$^{126m}$Sb (direct and daughter of $^{126}$Sn) &	-0.6(\textit{T})	&	250.9	&	185.9	\\  
$^{126}$Sb (direct and daughter of $^{126m}$Sb) &	-4.02(\textit{T})/0.08	&	167.6	&	126.4	\\  
\hline
\end{tabular}
\caption{Change in \% for the expected number of events when a different cross section or flux parametrization is used. The reference is the number of events obtained in one year, for one year exposure at sea level, with the ACTIVIA code \cite{activia} for \textit{E}$>$100\,MeV and, if available, the TENDL-2009 library \cite{talys} for 10\,MeV$<$\textit{E}$<$200\,MeV. See sec. \ref{sec::exc_funcion}.
\underline{Col.2}: \% difference in the expected number of events when a different routine is used to obtain the cross section values. In the high energy range (\textit{E}$>$100\,MeV) the YIELDX routine \cite{yieldx} has been used.  When available, the TENDL-2012 library has been used in the range 10\,MeV$<$ \textit{E} $<$ 200\,MeV. The difference due to the use of the TENDL-2012 library is given separately and identified with a \textit{T} in brackets.
\underline{Col.3}: \% difference in the expected number of events when the flux parametrization from Ziegler is used \cite{zie98}.
\underline{Col.4}: \% difference in the expected number of events when the flux parametrization from Gordon is used \cite{gor04}.
}
\label{tab::percent_change}
\end{table*}

\section{Excitation functions}\label{sec::exc_funcion}

\begin{figure}[h]\centering
\includegraphics[scale=0.48]{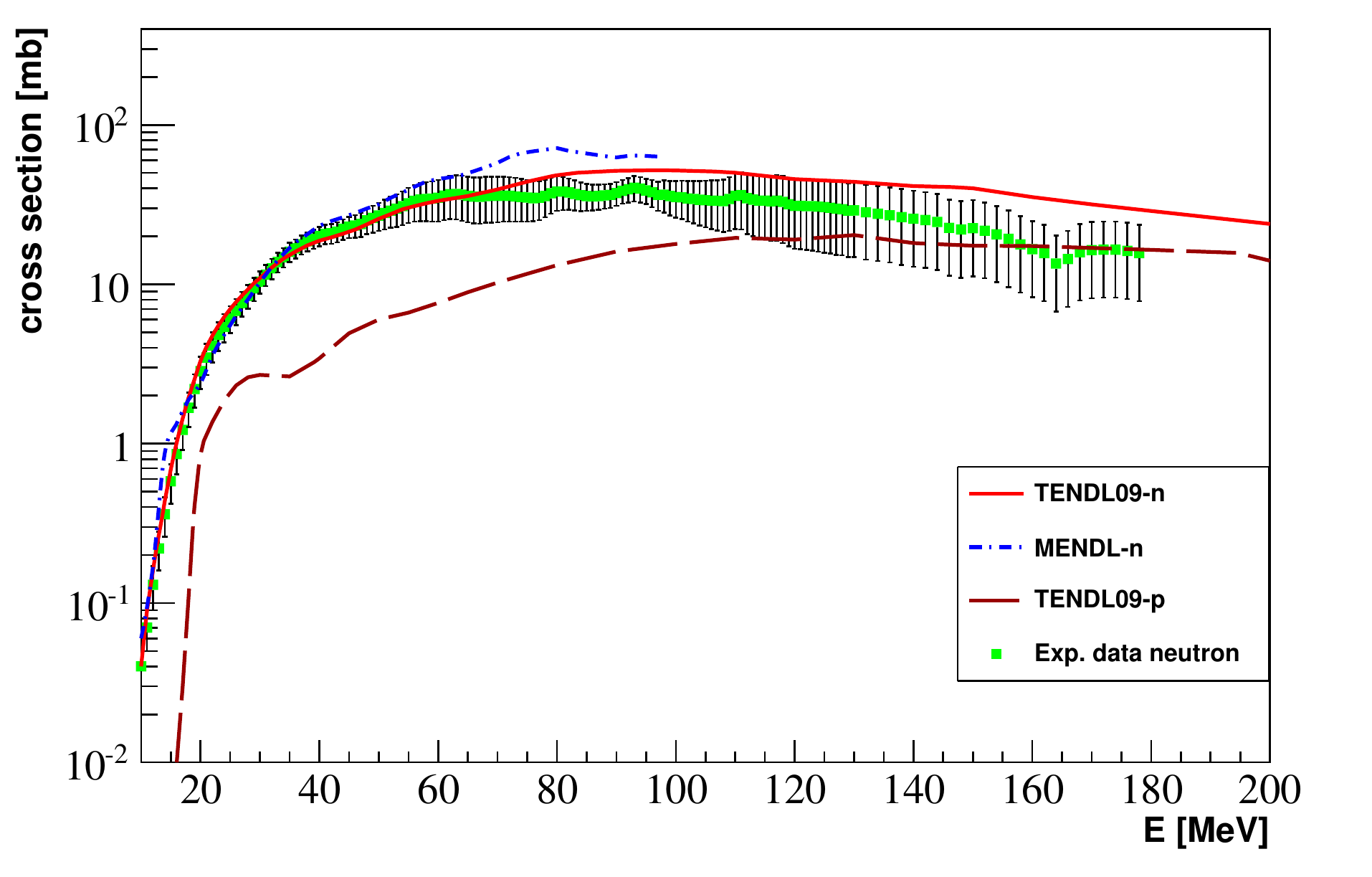}
\caption{Comparison for the $^{124}$Sb excitation function in the low energy range. Experimental data point are for the reaction \natte(n,x) \cite{hans10}. The calculated excitation function with the TENDL-2009 library \cite{talys} for neutrons (red, solid) shows a better agreement with the data than the result obtained from MENDL-n \cite{mendl} (blue, dashed dotted). The calculated excitation function for the \natte(p,x) reaction obtained from the TENDL-2009 proton library is also shown (red, dashed).}\label{fig::124Sb_low}
\end{figure}

The cross sections used for this evaluation are based on the semi-empirical formulas of Tsao and Silberberg \cite{silb73}\cite{silb77}\cite{yieldx}. The formulas are obtained for protons and valid for \textit{E}$>$100\,MeV. Even if obtained for protons, it is commonly considered that the neutron and proton cross sections are very similar at high energies. The formulas do not discriminate between ground and metastable state.\\
Semi-empirical cross section formulas have been integrated in several codes. An extensive comparison of available cross section data and simulation codes has been done by the IDEA collaboration (part of the ILIAS project) in order to find a set of standards for the germanium, copper and tellurium target materials \cite{Ceb10}\cite{IDEA}. The code used in the IDEA's evaluation is the YIELDX routine (by the authors of the semi-empirical formulas \cite{yieldx}) for energy larger than 100\,MeV and the MENDL library \cite{mendl} for the medium-low energy range.\\
In this analysis the ACTIVIA code \cite{activia} and the YIELDX routine are used for \textit{E}$>$100\,MeV. Both codes use the most up-to-date semi-empirical cross section formulas from 1998. Experimental data are also taken into account and used to correct the cross section formulas. The ACTIVIA code provides both the isotope cross section values and production rates for a natural target and a range of energies. The ACTIVIA code automatically includes the feeding from short lived parents. The YIELDX routine provides the cross section value at a fixed energy and for a particular product nuclide in a target nucleus. The YIELDX routine has been iterated to obtain the excitation functions and the production rates for products of a natural target. To compare the results with the ones obtained using the ACTIVIA code the excitation functions of the short lived parents have been calculated and included. The correct use of the YIELDX routine has been checked comparing the obtained cross section data at fixed energies (800\,MeV, 1400\,MeV and 23\,GeV) with the one in \cite{bar10}. The results obtained are the same but for the value at 800\,MeV of $^{110m}$Ag. However, it seems that the correct value for this isotope is mistyped in \cite{bar10}.\\
In the low energy range (\textit{E}$<$200\,MeV) neutron and proton cross sections can be very different. For this reason, and when available, the TENDL library for neutrons and protons \cite{talys} has been used for 10\,MeV$<$E$<$200\,MeV instead of the semi-empirical formulas. Semi-empirical formulas are then used above 200\,MeV. When the TENDL library is not available the semi-empirical formulas are used starting from 100\,MeV. The TENDL library has been used because it is updated yearly and shows a good agreement with the available experimental data. In figure \ref{fig::124Sb_low} the comparison among the MENDL, TENDL library and data for $^{124}$Sb in the low energy range is shown. The TENDL library shows a better agreement with the data.\\
In this paper the reference production rates are calculated using the TENDL-2009 library in the low energy range and the ACTIVIA code in the high energy range. The ACTIVIA code has the advantage to include the feeding chains for short lived parents and to calculate both the total cross section and production rate for a natural target.
In column 2 of table \ref{tab::percent_change} the obtained results (number of events for 1 year sea-level exposure and 1 year of measurement) are compared to the one calculated with the YIELDX routine (high energy range) and the TENDL-2012 library (low energy range). A $^{nat}$Te target was assumed (natural abundances from \cite{nudat2}). The use of the TENDL-2009 or TENDL-2012 library has a small effect on the expected number of events. The difference between the ACTIVIA code and the YIELDX routine can be due to the different implementation of the formulas of \cite{silb73}\cite{silb77} and especially of the later energy dependent updates \cite{yieldx}.\\
In the followings the excitation function obtained from the two codes has been compared with the available experimental data. Due to the lack of data for \textit{E}$>$100\,MeV only few isotopes can be compared to measurements.

\subsection{$^{46}$Sc}
\begin{figure}[h]\centering
\includegraphics[scale=0.48]{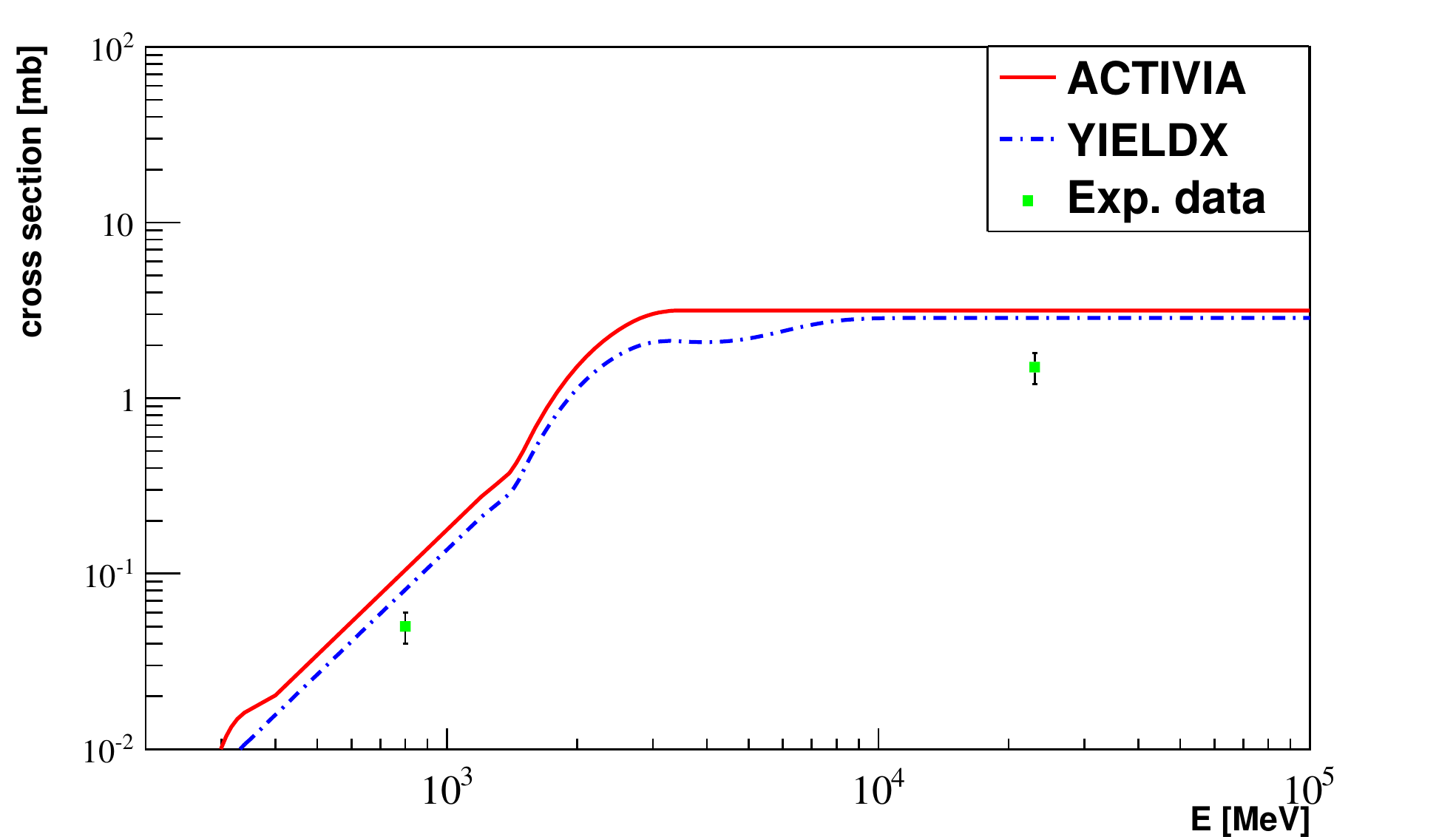}
\caption{Comparison for the $^{46}$Sc excitation function obtained with the ACTIVIA code (red, solid) and the YIELDX routine (blue, dashed dotted). Experimental data are from \cite{bar10}. The ACTIVIA and the YIELDX calculations show a good agreement. The theoretical excitation function predict a value that is a factor 2 larger than the experimental data.}
\label{fig::46Sc_xs}
\end{figure}

In figure \ref{fig::46Sc_xs} the comparison between the ACTIVIA and the YIELDX codes is shown together with the measured cross section data from \cite{bar10}. $^{46}$Sc decays in 99.996\% of the cases to the 2.01\,MeV excited state of $^{46}$Ti with an end-point energy of 0.36\,MeV (Q-value of 2.37\,MeV). It could be a concerning isotope for the neutrinoless double beta decay search in scintillator based experiments (wide region of interest) due to the decay in proximity of the expected signal. However, thanks to the relative short half-life, a time dependence analysis would make a discrimination with respect to the signal (constant in time) possible. It could also be efficiently reduced by an appropriate cooling down time.\\
The excitation functions calculated with the ACTIVIA code and the YIELDX routine agrees well. Both the codes predict a cross section value of a factor 2 higher than the experimental data.

\subsection{$^{58}$Co}
\begin{figure}[Ht]\centering
\includegraphics[scale=0.49]{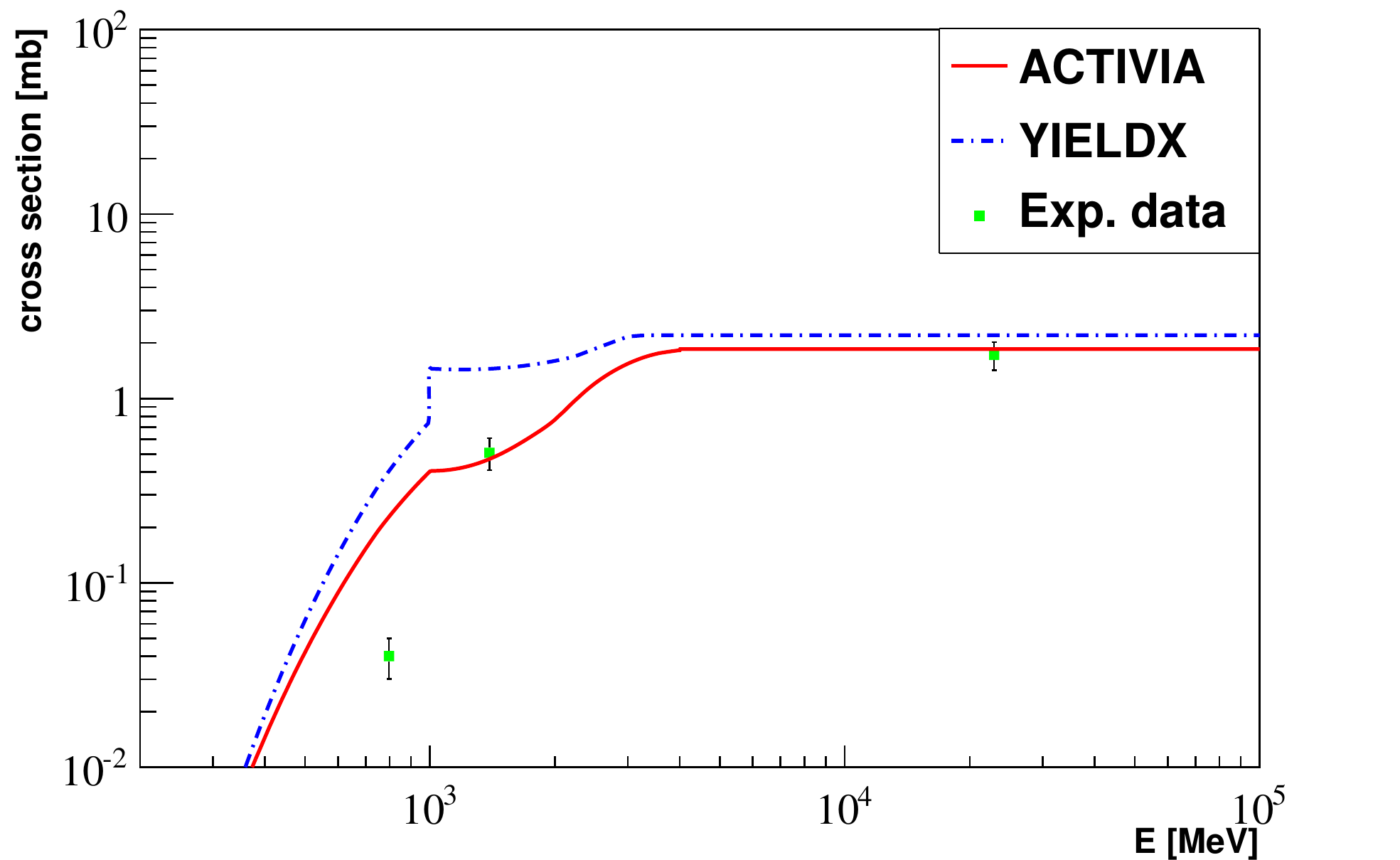}
\caption{Comparison for the $^{58}$Co excitation function obtained with the ACTIVIA code (red, solid) and the YIELDX routine (blue, dashed dotted). Experimental data are from \cite{bar10}. The ACTIVIA and the YIELDX calculations show a good agreement within 20\% at high energies. The calculated value agrees well with experimental data for energy above 800\,MeV.}
\label{fig::58Co_xs}
\end{figure}

In figure \ref{fig::58Co_xs} the comparison between the ACTIVIA and the YIELDX codes is shown together with the measured cross section data from \cite{bar10}. $^{58}$Co is not considered a big concern for the neutrinoless double beta decay search due to the relative short half-life and the decay mode. In 14.9\% of the cases it $\beta^{+}$-decays to the 0.81\,MeV excited state of $^{58}$Fe with an end-point energy of 0.47\,MeV (Q-value of 2.31\,MeV). Even with a 300\,keV energy resolution, only a small tail will fall in the energy region where the signal is expected.\\
The excitation function calculated with the ACTIVIA code and the YIELDX routine agrees well within 20\% at high energies. The ACTIVIA routine shows a better agreement with the measurement for energy above 800\,MeV. At low energy the experimental data point is a factor 5 lower than the calculation.

\subsection{$^{60}$Co}
\begin{figure}[t]\centering
\includegraphics[scale=0.48]{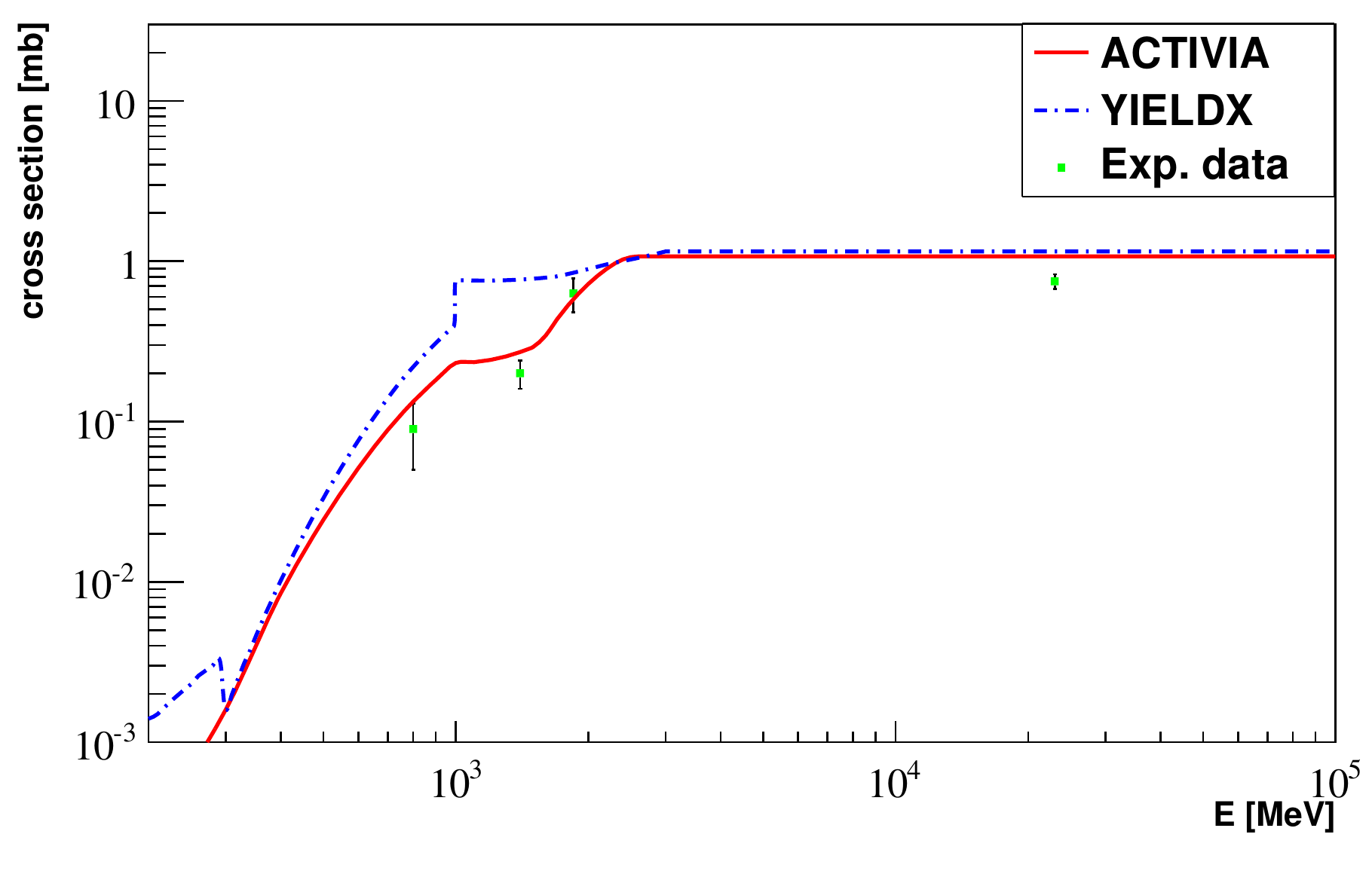}
\caption{Comparison for the $^{60}$Co excitation function obtained with the ACTIVIA code (red, solid) and the YIELDX routine (blue, dashed dotted). Experimental data are from \cite{bar10}\cite{nor05}. The ACTIVIA code agrees better to the experimental data.}
\label{fig::60Co_xs}
\end{figure}

$^{60}$Co is one of the most dangerous backgrounds for the search for neutrinoless double beta decay of $^{130}$Te both in bolometers and liquid scintillator based experiments. In the decay (Q-value of 2.82 MeV) two gammas (sum energy of 2.5 MeV) and one beta are emitted. The beta plus gamma spectral shape is very similar to the expected signal. The decay has been identified as a concern by the CUORE experiment. In addition, it has been one of the background candidates in KamLAND-Zen (scintillator based experiment for the search for neutrinoless double beta decay using $^{136}$Xe \cite{Kam12}). Xe and Te are close in mass number and it is expected that they have a similar behavior for what concerns neutron and proton activation (i.e. $^{60}$Co can be produced by spallation reactions both on Xe and Te). Additionally, the end-point energy of $^{136}$Xe is 2.46\,MeV, such that potential cosmogenic background candidates are in common for the two isotopes. It can be directly produced by spallation reactions on tellurium or born from the decay of $^{60}$Fe also produced by reaction on tellurium. In figure \ref{fig::60Co_xs} a comparison between the ACTIVIA and the YIELDX codes is shown together with the measured cross section data from \cite{bar10}\cite{nor05}. The ACTIVIA code shows a better agreement with the experimental data. The difference between the ACTIVIA and the YIELDX code could be due to a different tuning of the excitation functions to match the experimental data. This fact was already noticed in \cite{IDEA}, where the excitation function from YIELDX is compared to the one obtained with a modified version of COSMO \cite{cosmo}. The COSMO code is also based on the semi-empirical formulas \cite{silb73}\cite{silb77}. However, the excitation function obtained with the COSMO code in \cite{IDEA} was one order of magnitude higher in the energy region between 1\,GeV and 3\,GeV, and a factor 2 above 3\,GeV.\\
For the ACTIVIA and the YIELDX code the disagreement is mainly between 1\,GeV and 2\,GeV, where the excitation function obtained with the YIELDX routine seems to show an artefact. The matching with the experimental data could be improved with further tuning of the parameters in the codes, as suggested in \cite{IDEA}.

\subsection{$^{84}$Rb}
\begin{figure}[h]\centering
\includegraphics[scale=0.48]{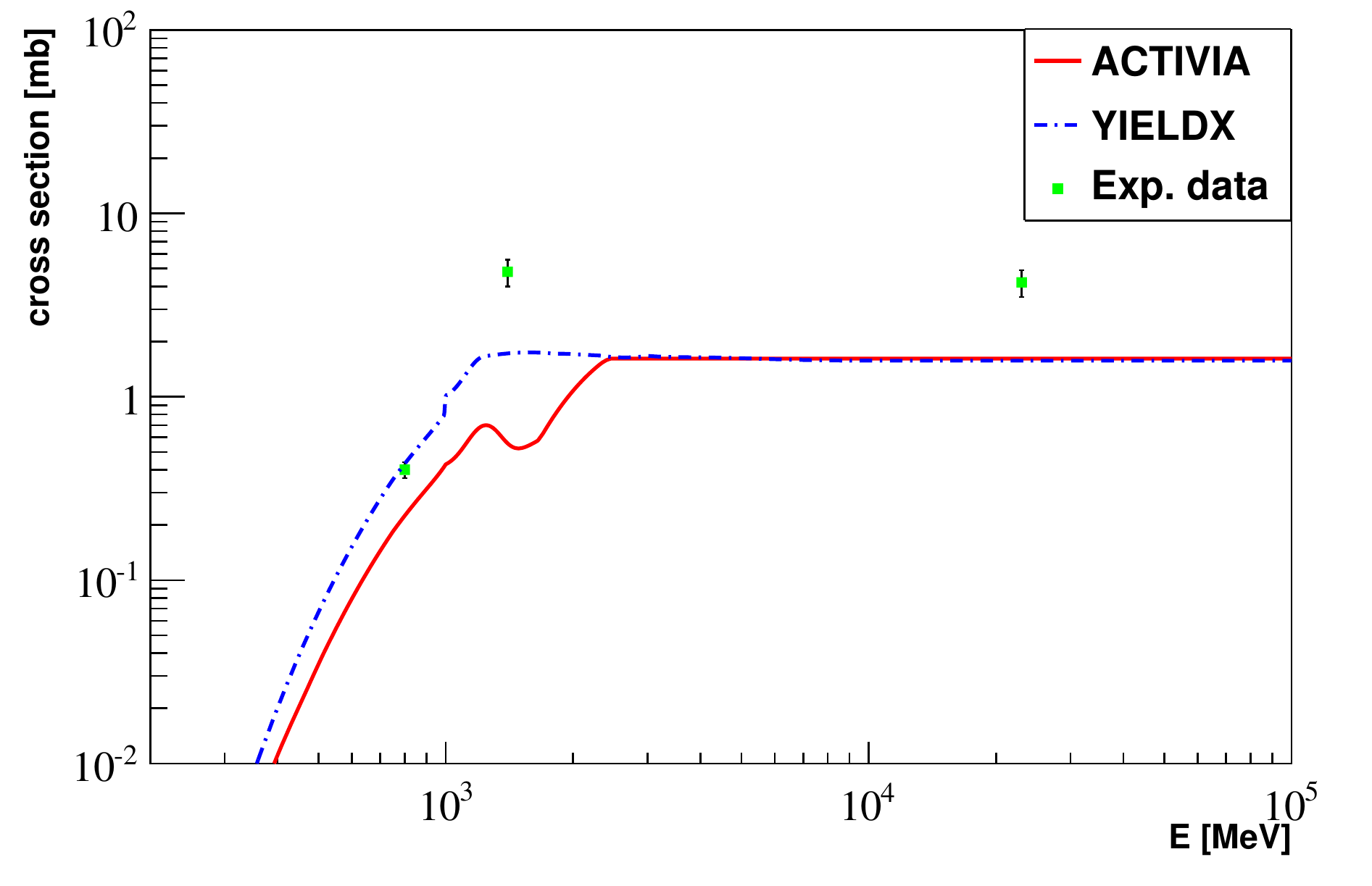}
\caption{Comparison for the $^{84}$Rb excitation function obtained with the ACTIVIA code (red, solid) and the YIELDX routine (blue, dashed dotted). Experimental data are from \cite{bar10}. The ACTIVIA and the YIELDX calculations show a good agreement above 2\,GeV. The experimental data are a factor 2 higher than the calculated value for energy above 1\,GeV.}
\label{fig::84Rb_xs}
\end{figure}

In figure \ref{fig::84Rb_xs} the comparison between the ACTIVIA and the YIELDX codes is shown together with the measured cross section data from \cite{bar10}. $^{84}$Rb is not considered a nuclide of concern due to the short half-life as noted for $^{58}$Co.\\
The excitation function calculated with the ACTIVIA code and the YIELDX routine agrees well above 2\,GeV. The disagreement below 2\,GeV could be a different implementation of the cross section parametrization. The YIELDX routine shows a better agreement with the measurement at 800\,MeV. At high energy the experimental data point is a factor 2 higher than the calculation.

\subsection{$^{88}$Y}
\begin{figure}[t]\centering
\includegraphics[scale=0.49]{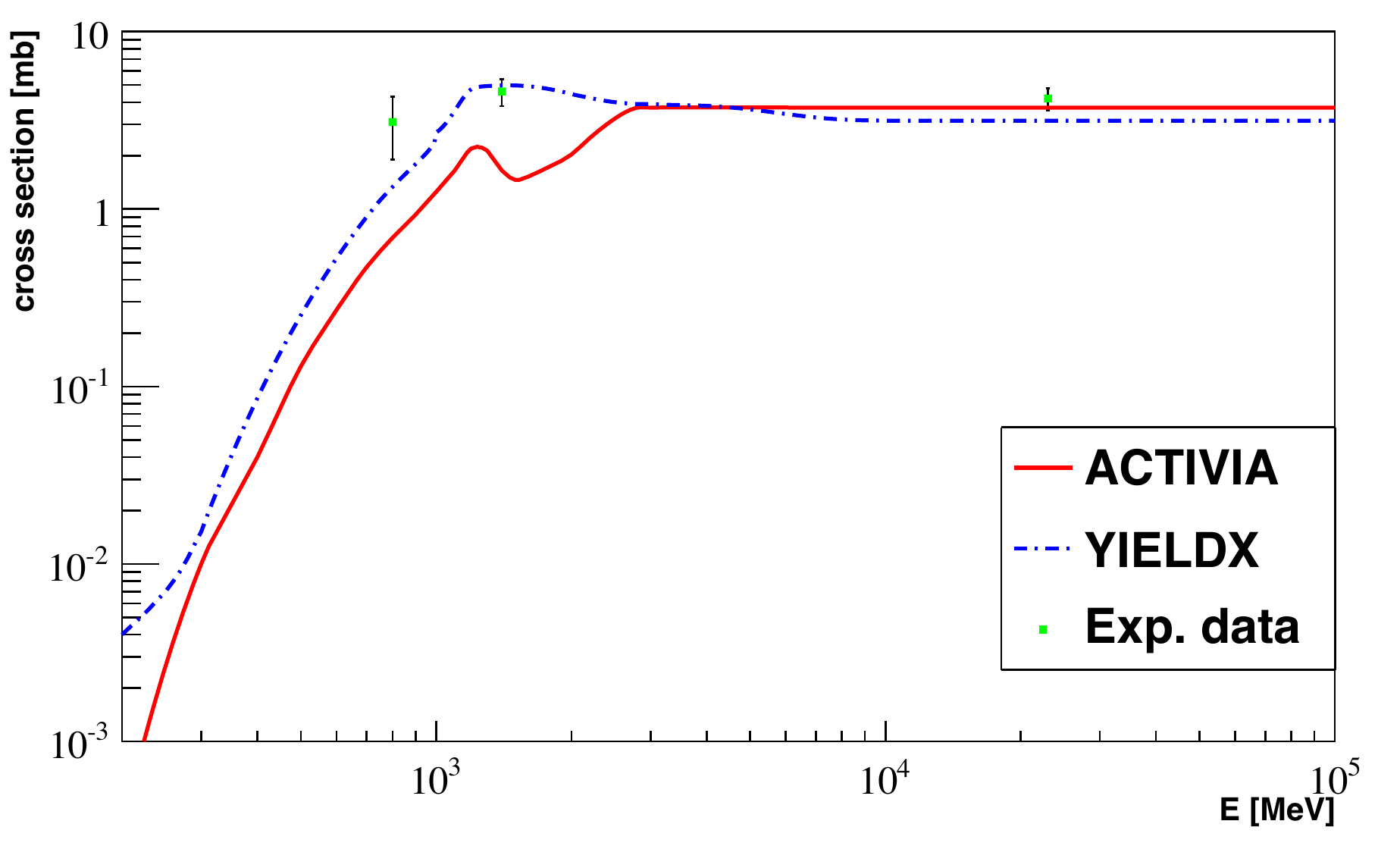}
\caption{Comparison for the $^{88}$Y excitation function obtained with the ACTIVIA code (red, solid) and the YIELDX routine (blue, dashed dotted). Experimental data are from \cite{bar10}. The values from the ACTIVIA code show a better agreement with the experimental data for high energies, while the YIELDX routine shows a better agreement at low energies.}
\label{fig::88Y_xs}
\end{figure}

$^{88}$Y has been identified as a potential cosmogenic background candidate for the KamLAND-Zen experiment (scintillator based experiment). It can be directly produced by spallation reactions on tellurium and fed by the decay of $^{88}$Zr, also produced in spallation reactions. The isotope electron captures to the 2.7\,MeV excited state of $^{88}$Sr in 94.5\% of the cases.\\
The comparison for the $^{88}$Y excitation function is shown in figure~\ref{fig::88Y_xs}. In this case the YIELDX routine shows a better agreement with the data in the low energy range (\textit{E}$<$1\,GeV). The ACTIVIA codes agrees better with the data for \textit{E}$>$2\,GeV. A larger deviation between the two codes is visible between 1\,GeV and 2\,GeV. The difference could be due to a different tuning of the semi-empirical formulas in the codes. Deviation factors between experimental data and values obtained with the ACTIVIA code are 4.5, 2.8 and 1.1 for the 800 MeV, 1.4 GeV and 23 GeV energies respectively.

\subsection{$^{102}$Rh}
\begin{figure}[h]\centering
\includegraphics[scale=0.49]{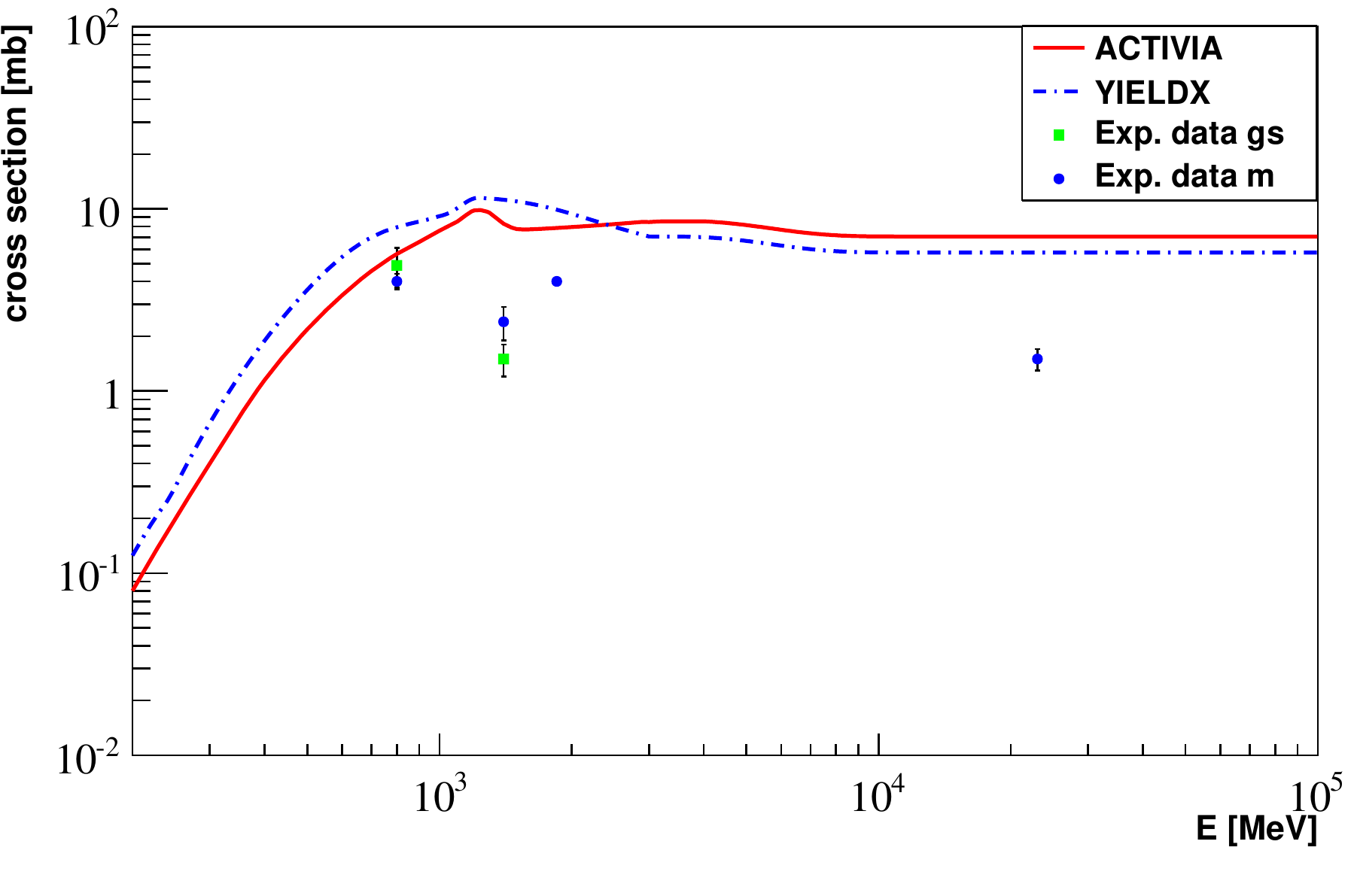}
\caption{Comparison for the $^{102}$Rh excitation function obtained with the ACTIVIA code (red, solid) and the YIELDX routine (blue, dashed dotted). Experimental data are from \cite{bar10}\cite{bar97}. The ACTIVIA and the YIELDX calculations show a discrepancy of 20\% but have a similar shape. The experimental data are a factor 5 lower than the calculated values for energy above 1\,GeV.}
\label{fig::102Rh_xs}
\end{figure}

$^{102}$Rh, ground and metastable state, are considered of a minor concern for the search for neutrinoless double beta decay of $^{130}$Te. The highest excited states of $^{102}$Ru reached by electron capture is 2.22\,MeV (2.26\,MeV) for the $^{102}$Rh-m ($^{102}$Rh-gs) decay. Only the tail of the decay (and only for scintillator based experiments) will fall in the energy region where the signal is expected.\\
In figure \ref{fig::102Rh_xs} the comparison between the ACTIVIA and the YIELDX codes is shown together with the measured cross section data from \cite{bar10}\cite{bar97} for the ground (gs, green, square) and excited (m, blue, circle) states. The excitation function calculated by the ACTIVIA code and the YIELDX routine agrees within 20\%. The calculation does not distinguish between the ground and the excited state. The available experimental data suggests that the cross section of the ground and the excited state are of the same order of magnitude. This suggests that the same excitation function can be used for the two states. For energy above 1\,GeV the calculated excitation functions are a factor 5 higher than the experimental data. 

\subsection{$^{110m}$Ag}
\begin{figure}[ht]\centering
\includegraphics[scale=0.48]{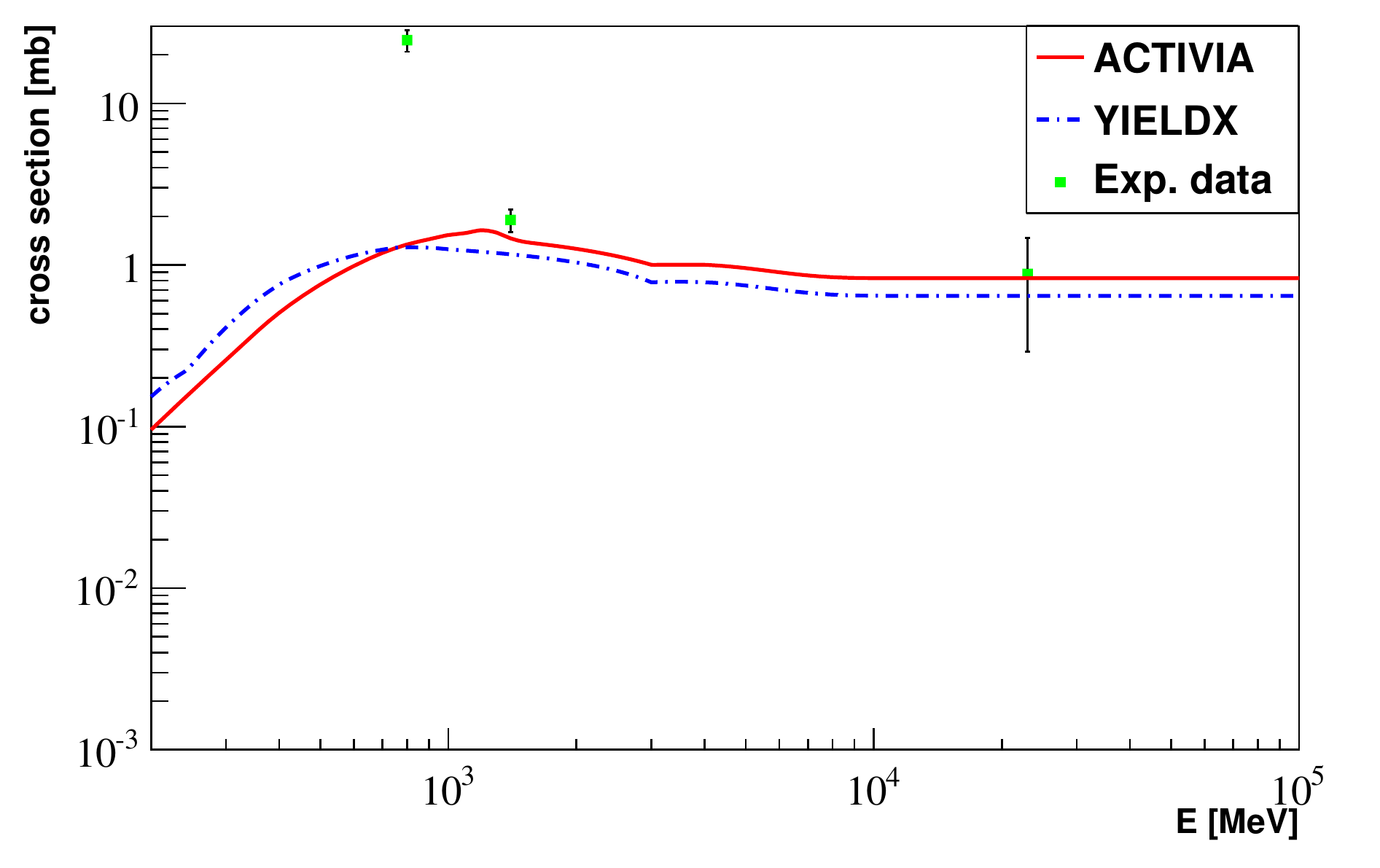}
\caption{Comparison for the $^{110m}$Ag excitation function obtained with the ACTIVIA code (red, solid) and the YIELDX routine (blue, dashed dotted). Experimental data are from \cite{bar10}.}
\label{fig::110Agm_xs}
\end{figure}

$^{110m}$Ag has been identified has a potential background candidate by the KamLAND-Zen experiment. It has also been considered a worrying background by the CUORE collaboration. In 30.8\% of the cases it $\beta^{-}$-decays to the 2.48\,MeV excited state of $^{110}$Cd with a Q-value of 3.01\,MeV. The decay is where the signal is searched for.\\
The comparison for the $^{110m}$Ag excitation function is shown in figure \ref{fig::110Agm_xs}.  The calculation with ACTIVIA code and the YIELDX routine exhibits a good agreement. Above 800 MeV, the cross section was measured for two proton energies (1.4\,GeV and 23\,GeV) \cite{bar10}. The agreement with the ACTIVIA code is fairly good (deviation factor of 1.3 for 1.4 GeV and 1.06 for 23 GeV energies \cite{IDEA}). The experimental data point at 800 MeV \cite{bar10} is about a factor 20 higher than the value predicted by the code. This value seems suspiciously too high.\\
This is supported by a recent measurement at the LANSCE facility \cite{wan12} in the energy range from 1.25 MeV to 800 MeV. The preliminary value for the averaged cross section is 0.18$\pm$0.03\,mb over the entire energy range. The result seems to show a better agreement with the semi-empirical formulas than the value at 800\,MeV in \cite{bar10}.

\subsection{$^{124}$Sb}
\begin{figure}[h]\centering
\includegraphics[scale=0.48]{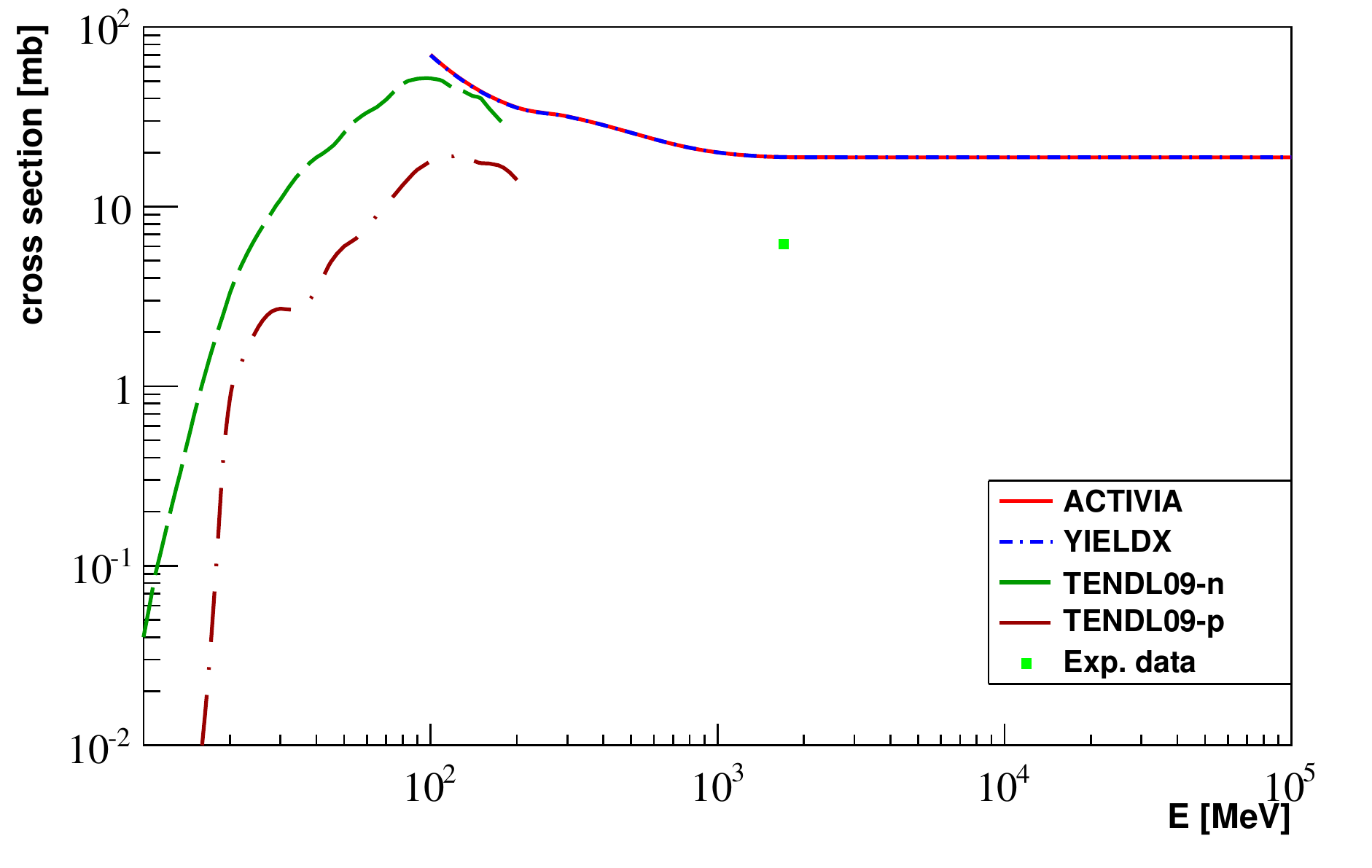}
\caption{Comparison for the $^{124}$Sb excitation function obtained with the ACTIVIA code (red, solid) and the YIELDX routine (blue, dashed dotted). Experimental data are from \cite{por00}. The ACTIVIA and the YIELDX calculation show a perfect agreement. In the low energy range the calculation obtained with the TENDL-2009 neutron (green, long dashed) and proton (brown, long dashed dotted) library are shown. They present a good agreement with the high energy range excitation function.}
\label{fig::124Sb_xs}
\end{figure}

$^{124}$Sb, despite the relative short half-life can be a direct background for the neutrinoless double beta decay search in tellurium (both bolometers and scintillator based experiments). It has a high production cross section at low energies. In order to reduce the produced activity during exposure at sea-level to a negligible amount, long cooling down times and/or a purification process are necessary. It has a complex decay scheme with multiple gamma and beta emitted. In 8.8\% of the cases it decays to the 2.69\,MeV excited state of $^{124}$Te with a Q-value of 2.90\,MeV. The relative flat spectrum in the region where the signal is searched for makes it a potential dangerous candidate as already noted in \cite{IDEA}.\\ 
In figure \ref{fig::124Sb_xs} the comparison between the ACTIVIA and the YIELDX codes is shown together with the measured cross section data from \cite{por00}. The excitation function calculated by the ACTIVIA code and the YIELDX routine shows a perfect agreement. However, they are both a factor 3 higher than the experimental data at 1.7\,GeV. 

\subsection{Other decays}

For the other isotopes listed in table \ref{tab::isotopes} there are not found experimental data to which compare the calculated excitation function. Among them the ones considered most dangerous based on their decay mode are $^{22}$Na, $^{42}$K, $^{44}$Sc, $^{68}$Ga and $^{106}$Rh.\\
$^{22}$Na $\beta^{+}$-decays to the 1.27\,MeV excited state of $^{22}$Ne. The decay is mainly dangerous for scintillator based experiment, where the 511\,keV are summed to the other gamma emitted and not distinguished from them. This effect will create a beta spectrum from 2.29\,MeV to 2.84\,MeV which can mimic the expected signal.\\
$^{42}$K and $^{106}$Rh have a very short half-life but are fed by the decay of the long-lived $^{42}$Ar and $^{106}$Ru, respectively. They both present a flat spectrum in the energy region where the neutrinoless double beta decay is searched for.\\
$^{44}$Sc electron captures to the 2.66\,MeV excited state of $^{44}$Ca in 1.02\% of the cases. Like in the case of $^{42}$K, it is relative short lived but is fed by the decay of the long-lived $^{44}$Ti.\\
Finally, $^{68}$Ga is considered a medium level concern. It electron captures to the 2.34\,MeV excited state of $^{68}$Zn in 0.09\% of the cases. In addition, in scintillator based experiments, the tail of the $\beta^{+}$-decay will leak in the region where the signal is searched for. It is then dangerous when present in high abundance. It is also short lived but fed by the long-lived $^{68}$Ge.\\
$^{26}$Al and $^{126}$Sn (that decays to the short lived but high Q-value $^{126m}$Sb) have extremely long T$_{1/2}$. They are considered not dangerous unless the tellurium spent long time (several tens of years) at sea level before being used for neutrinoless double beta decay studies.

\section{Cosmogenic-induced background}

The expected production rates are calculated using the ACTIVIA code for the isotopes listed in table \ref{tab::isotopes}. Corrections to the expected activity are applied when a nuclide is directly produced by cosmogenic spallation reactions and simultaneously formed by the decay of another radionuclide born in the target. This is for instance the case for $^{60}$Co, which can be produced by cosmogenic neutron and proton bombardment on a $^{nat}$Te target or by the decay of $^{60}$Fe. The formulas used to calculate the expected activity are the following \cite{Leb12}:
\begin{equation}
\begin{split}
A_{1}^{EOB}&=R_{1}(1-e^{-\lambda_{1}t_{exp}}),\\
A_{1}(t)&=A_{1}^{EOB}e^{-\lambda_{1}t},\\
A_{2}^{EOB}&=R_{2}(1-e^{-\lambda_{2}t_{exp}})+\\
&+R_{1}\cdot f\left( 1- \frac{\lambda_{2}}{\lambda_{2}-\lambda_{1}}e^{-\lambda_{1}t_{exp}}+\frac{\lambda_{1}}{\lambda_{2}-\lambda_{1}}e^{-\lambda_{2}t_{exp}} \right),\\
A_{2}(t)&=A_{2}^{EOB}e^{-\lambda_{2}t}+A_{1}^{EOB}\cdot f \cdot \frac{\lambda_{2} } {\lambda_{2}-\lambda_{1}}(e^{-\lambda_{1}t} -e^{-\lambda_{2}t}),
\end{split}
\end{equation}
where 1 and 2 are the index for the parent and daughter nuclide respectively. \textit{A$^{EOB}$} is the activity at the end of bombardment (\textit{t$_{exp}$}), \textit{f} is the transition probability of a parent to a daughter nuclide and \textit{R} is the production rate.
Results are shown in table \ref{tab::Te_cosmo_1}. The expected activity ($\mu$Bq/kg of natural tellurium) has been calculated assuming one year of surface exposure and two different cooling down scenarios:
\begin{enumerate}
\item SNO+ like scenario.  A first purification stage on surface and additional 6 months of cooling down time deep underground. The purification factor (PF) for the telluric acid has been assumed to be 10$^{-4}$ \cite{Bil13}. This factor can be achieved in a two step telluric acid re-crystallization process developed by the SNO+ collaboration at BNL \cite{yeh13}. The purification has not only the aim of reducing the potential cosmogenic background but also to reduce the U and Th backgrounds. The purification factor has been tested over a wide range of isotopes adding spikes to the Te-loaded scintillator and then assaying the purified sample with X-ray fluorescence analysis. After the surface purification an additional exposure of 5 hours has been assumed to account for the time needed to transport the material deep underground; 
\item CUORE like scenario. A very high purity powder is used for the grown of the TeO$_{2}$ crystals on surface \cite{Arn10}. Once prepared the crystals are shipped to Italy where they are stored underground. Two years of cooling down deep underground has been assumed in this case.
\end{enumerate}
The expected number of events in one year of counting per tonne of material is also shown in table \ref{tab::Te_cosmo_1} for the two cooling scenarios. Decay branching ratios are taken into account. \\
In case of the CUORE experiment, thanks to the good energy resolution (5\,keV FWHM) and the long cooling down underground, the most dangerous backgrounds are expected to be $^{110m}$Ag and $^{60}$Co \cite{bar10}. The other isotopes have a maximum energy deposited in the single crystal far below (or above) the 2.5\,MeV and thus would not enter in the ROI. With a two years cooling down time the reached activities are within the values given in \cite{CUORE08} (0.2 $\mu$Bq/kg of TeO$_{2}$ for $^{60}$Co). In addition, the activities will be further reduced by a shorter exposure time at sea-level (4 months instead of the assumed one year).\\
The conclusion is different in case of the SNO+ experiment. The expected energy resolution is much worse than in case of CUORE. It must be also considered the fact that the total energy released in the decay is measured. With a total volume of 780\,t, the emitted gammas and beta deposit their total energy in the detector and are seen as a single event. For example, in the $\beta^{+}$-decay the energy of the two 511\,keV gammas (annihilation gammas) has to be added to the energy of the positron. The result is that the beta spectrum is shifted of 1.022\,MeV. Following this consideration, the number of isotopes that could be a potential background increases, as discussed in section \ref{sec::exc_funcion}. The aim of the SNO+ collaboration is to reduce the total cosmogenic-induced background to a negligible level, few counts/yr, in the region of interest.
Based on the results of table \ref{tab::Te_cosmo_1}, and considering the decay mode of the diverse isotopes, it can be concluded that even after two years of cooling down time following one year of exposure at sea-level, $^{60}$Co, $^{110m}$Ag, $^{22}$Na and $^{88}$Y still pose a problem. It must also be noted, that it is not known yet if such long cooling down times are possible. For a shorter cooling down time underground the number of potential background candidates will increase. The other scenario investigated, where the tellurium acid is purified after exposure at sea-level and before being loaded in the detector, shows a larger reduction for the cited nuclides already at the end of the purification process. Assuming a further 6 months of cooling underground to remove the short lived isotopes that could have formed after purification, reduces the potential backgrounds to $^{124}$Sb, $^{88}$Y and $^{110m}$Ag. The expected activity is at least one order of magnitude smaller than in the previous case. Of these, $^{124}$Sb will decay quickly with time.

\subsection{$^{60}$Co}
The production rate for $^{60}$Co on $^{nat}$Te has been widely investigated being one of the most dangerous nuclide for the tellurium \zn~ search. To cross check and support the results obtained in this paper, the value is compared to previous calculations \cite{activia}\cite{IDEA}\cite{bar10}.\\
In the energy range from 10 MeV to 100 GeV (10 MeV energy step) the calculated production rate is 0.81 $\mu$Bq/kg. This is the same value obtained in \cite{activia}.\\
The activity calculated with the experimental data in \cite{bar10} for 45 days of exposure agrees within of a factor 2 (the difference could be due to the cosmic spectrum binning). \\
Finally the production rate obtained by the IDEA group for the 10 MeV to 10 GeV energy range using the flux parametrization from Armstrong and the YIELDX routine (0.7 $\mu$Bq/kg of TeO$_{2}$ \cite{IDEA}), agrees well with the value calculated in this paper using the same routine.

\begin{table*}[ht]\centering
\scalebox{0.97}{
\begin{tabular}{cccccccc}
  \hline
Isotope 	&	 \textit{R}  ($\phi$ from \cite{arm73}\cite{geh85}) & \multicolumn{2}{c}{Events/t in 1 yr} & A [$\mu$Bq/kg] & Events/t in 1 yr & A [$\mu$Bq/kg]& Events/t in 1 yr \\ \cline{3-4} \cline{5-6}\cline{7-8}
	&	[$\mu$Bq/kg]	&	\textit{t$_{exp}$}=1 yr & PF = 10$^{-4}$+5h & \multicolumn{2}{c}{\textit{t$_{cool}$}=6 months} & \multicolumn{2}{c}{\textit{t$_{cool}$}=2 yrs}\\\hline
$^{22}$Na 	&	 1.01 	&	6.54E+3	&	4.90	&	1.55E-4 & 4.29	&	0.14 & 3.84E+3\\  	
$^{26}$Al 	&	0.67	&	0.02	&	1.37E-5	&	4.34E-10 & 1.37E-5	&	6.48E-7 & 0.02\\  
$^{42}$K 		&	1.33 (0.24)	&	85.11+156.25	&	20.87+0.11	&	(3.35E-6) & 0.10	& (0.005) & 149.806\\  
$^{44}$Sc 		& 1.19 (0.052)	&	24.54+19.02	&	14.29+0.01	&	(4.06E-7) & 0.01	&(5.92E-4) &	18.58\\  
$^{46}$Sc 	&	1.97	&	1.86E+4	&	35.56	&	7.90E-4 & 7.85	&0.004 &	44.21\\  
$^{56}$Co 	&		0.13	&		1.12E+3	&	2.29	&	4.81E-5 & 0.45	&1.72E-4 &	1.60 \\  
$^{58}$Co 	&	 1.29	&	1.08E+4	&	23.62	&	4.61E-4 &3.96	&	9.91E-4 &8.51 \\  
$^{60}$Co 	&	 0.81 (0.367) &	2.95E+3	&	2.09	&	6.62E-5 & 1.96	&	0.077 &2.27E+3 \\  
$^{68}$Ga 	&			3.14 (1.28)	&	21.17+1.59E+4	&	17.55+15.58	&	(4.76E-4) &9.77	&(0.120) &	2.46E+3 \\  
$^{82}$Rb 	&		(2.44)	&	7.71E+3	&	44.58	&	(9.57E-5) &0.30	&(5.16E-9 )&	1.63E-5 \\  
$^{84}$Rb &		 1.29 	&	5.06E+3	&	22.76	&	1.17E-4 &0.48	&	2.45E-7 & 1.00E-3 \\  
$^{88}$Y	&		3.14 (8.11)	&	1.67E+5	&	176.68	&	0.006(0.003) & 99.05	&0.253(0.018) &	3.19E+3  \\  
$^{90}$Y	&		 0.69 (0.165) 	& 229.22+122.35	&	12.10+0.08	&	(2.63E-6) & 0.08	&(0.004) &	116.63 \\  
$^{102}$Rh &		 11.77 (0.03) 	&	1.18E+5	&	128.31	&	0.004 & 69.70	&	0.566 & 1.03E+4 \\  
$^{102m}$Rh &		11.77 	&	5.72E+4	&	41.46	&	0.001 & 37.79	&	1.371 & 3.95E+4 \\  
$^{106}$Rh	&		(0.06)	&	655.58	&	0.58	&	(1.81E-5) & 0.41	&(0.007) &	167.948 	 \\  
$^{110m}$Ag &	 2.39&	2.98E+4	&	29.97	&	9.10E-4 & 18.06	&	0.198 & 3.92E+3 \\  
$^{110}$Ag	&		(0.03)	&	401.17	&	0.40	&	1.22E-5 & 0.24	&0.003 &	52.86 \\  
$^{124}$Sb 	&	182.0	&		1.33E+6	&	3.36E+3	&	0.055 & 409.77	&	0.040 & 294.741 \\  
$^{126m}$Sb &		71.42 (7.91)	&	102.46	&	101.81	&	(1.37E-8) & 4.32E-4 &	(2.04E-5) &	0.64\\  
$^{126}$Sb	&		89.65 ($^{126m}$Sb)	& 1.53E+5	&	1.80E+3	&	4.13E-5 & 0.06	&(3.32E-6) &	0.10 \\  
\hline
\end{tabular}}
\caption{Expected production rate (\textit{R}) estimation for the isotopes listed in table \ref{tab::isotopes}. The values are obtained using the ACTIVIA code \cite{activia} for \textit{E}$>$100\,MeV (energy step of 10\,MeV). If available the cross sections for 10\,MeV$<$\textit{E}$<$200\,MeV are obtained from the TENDL-2009 library \cite{talys} (energy step of 10 eV). The flux parametrization is the one from Armstrong and Gehrels \cite{arm73}\cite{geh85}. Short and long-lived feeding parents have been included. Long-lived parent isotope production rates and activities are given in brackets. For very short lived isotopes (seconds) fed by a long-lived parent secular equilibrium has been assumed. When short lived (minutes to hours) isotopes are fed by long-lived ones the two contributions are shown separately. No correction factors are applied for the location on surface. Expected numbers of events are given for one year and per tonne (t) of material (natural tellurium). The decay branching ratios are taken into account. The two different scenarios are outlined in the text.}
\label{tab::Te_cosmo_1}
\end{table*}

\section{Underground activation}
In order to reduce the cosmogenic-induced background, it is important to minimize the exposure time of the material on surface and to maximize the storage underground. In addition, purification techniques could help reducing the produced background isotopes.\\
During the period the material is stored underground, it could be activated by the local neutron flux. At a depth of a few tens of meter water equivalent, the nucleonic component of the cosmic flux is reduced to a negligible level. The main sources of neutrons are low in energy (thermal neutrons and neutrons from ($\alpha$,n) reactions in the surrounding rock). It is expected that the underground activation has a smaller impact with respect to the one on surface. Nevertheless, it is important to estimate the activation of tellurium once underground, specially for long cooling down times when the isotopes produced on surface will decay away, leaving as major contribution the activity produce during the storage. To reduce the activation underground it is possible to use a water shielding around the detector. Materials with high hydrogen content will stop the low energy neutrons.\\
In the particular case of the SNO+ experiment, tellurium will be stored at SNOLAB underground lab, at a depth of about 2 km (6000 m.w.e). At this depth the main contributions to the underground neutron flux are:
\begin{itemize}
\item Neutrons from ($\alpha$,n) reaction in the rock. The total flux has been estimated to be around 4000 n/(m$^{2}\cdot$d) \cite{hea88}\cite{hea90}\cite{hea89} for norite rock. The neutron energy is below 15 MeV and peaked around 2.5 MeV \cite{vaz13};
\item Neutrons induced by muon interactions in the rock. The muon induced neutron flux is from \cite{mei05} and has an energy up to 3.5 GeV. The integrated flux is 5.4$\cdot 10^{-11}$\,n/(s$\cdot$cm$^{2}$). This flux is about a factor 60 smaller than the one at LNGS, where CUORE is located;
\item Thermal neutrons from the rock. The total thermal neutron flux is about 4145\,n/(m$^{2}\cdot$d) \cite{bro10}. The isotopes produced by thermal neutron activation on $^{nat}$Te are low in Q-value and/or short lived. 
\end{itemize}

The presence of a water shielding around the detector, like in the case of SNO+, reduces the thermal and fast neutron flux during data taking. 
For the isotopes shown in table \ref{tab::isotopes}, the underground activation during storage has been estimated to be less than 1 event/(yr$\cdot$t) at the SNOLAB depth.\\
For the isotopes listed in table \ref{tab::lowQ}, the production rates both at sea level and underground are shown in table \ref{tab::short_R}. The activation due to the thermal neutrons at sea-level has been included. For one year of surface exposure and 6 months of cooling down underground the major contribution comes from $^{127m}$Te with a rate of about 0.25 Bq/t.\\
An important isotope that can be produced during data taking by thermal neutron capture is $^{131m}$Te (Q-value=2.41\,MeV, T$_{1/2}$=33.25 h). It is produced by thermal neutron capture on $^{130}$Te. Due to the short half-life only the isotopes produced during storage underground will contribute to the initial data taking phase. The activation during data taking can be reduced with the help of a water shielding like in the SNO+ experiment.\\ 
An additional source of neutrons during data taking are the ($\alpha$,n) reactions on the atoms of the liquid scintillator. The ratio of thermal neutron capture on $^{nat}$Te with respect to the one on protons for the case of the SNO+ experiment (0.3\% loading), is 0.28\%. A small fraction of this (about 1.4\%) is on $^{130}$Te.

\begin{table*}[ht]\centering
\begin{tabular}{cccc}
   \hline
Isotope	& \textit{R}$_{surf}$ [$\mu$Bq/kg]& \textit{R}$_{\alpha,n}$ [$\mu$Bq/kg] & \textit{R}$_{rock}$ [$\mu$Bq/kg]\\\hline
$^{121m}$Sn & 29.71 & 3.20E-7 & 2.56E-7 \\ 
$^{123}$Sn & 26.88 & 7.33E-7 & 2.39E-7 \\ 
$^{125}$Sn & 34.01 & 7.02E-8 &  1.25E-7  \\ 
$^{120m}$Sb & 159.13 & 3.84E-6 & 1.25E-6   \\ 
$^{125}$Sb & 205.59 & 3.46E-5 &  2.36E-6  \\ 
$^{127}$Sb & 165.46 & 4.66E-12  & 2.12E-6 \\ 
$^{129}$Sb & 83.83 & 1.05E-12  &  1.36E-6  \\ 
$^{118}$Te & 203.02 & - &  1.11E-6\\ 
$^{121}$Te & 87.83 & 3.32E-3 & 1.41E-6\\ 
$^{121m}$Te & 376.27 & 9.16E-4 & 3.39E-6\\ 
$^{123m}$Te & 464.15 & 0.096 & 8.65E-6 \\ 
$^{125m}$Te & 791.04 & 0.349 & 1.42E-5\\ 
$^{127m}$Te & 680.83 & 5.30E-2 +$^{127}$Sb & 1.72E-5+$^{127}$Sb \\ 
$^{129m}$Te & 549.059 & 1.89E-2 & 1.33E-5 \\ 
$^{125}$I & 50.14  & -  & -\\ 
$^{126}$I & 35.76 & -  & -\\ 
$^{129}$I & 16.16 & - & -\\ 
$^{48}$V & 0.796 & - & -\\ 
$^{106m}$Ag & 23.191 & - & -\\ 
\hline
\end{tabular}
\caption{Production rates for isotopes listed in table \ref{tab::lowQ}. \textit{R$_{surf}$} is the production rate at sea level obtained with the ACTIVIA code for \textit{E}$>$200\,MeV and the TENDL-2009 library for \textit{E}$<$200\,MeV. \textit{R$_{\alpha,n}$} is the production rate at the SNOLAB location due to neutrons from ($\alpha$,n) reactions in the norite rock after 50 cm attenuation \cite{vaz13}. Cross sections values are from the TENDL-2009 library. \textit{R$_{rock}$} is the production rate due to $\mu$-induced neutron flux at the SNOLAB location. Values are obtained using the ACTIVIA code for \textit{E}$>$200\,MeV and the TENDL-2009 library for \textit{E}$<$200\,MeV. The flux is from \cite{mei05}. Unless otherwise specified the feeding isotope activities are included.}
\label{tab::short_R}
\end{table*} 

\section{Conclusion}
In this article the production rate for cosmogenic-induced isotopes on a natural tellurium target has been calculated. An extensive set of isotopes with Q-value larger than 2 MeV and T$_{1/2}$ larger than 20 days has been considered as potential background candidates. In addition, the production rates of shorter lived and low Q-value isotopes close in atomic mass to tellurium have been studied. High event rates can produce pile-up background. Production rates are obtained with the help of the ACTIVIA code for \textit{E}$>$100\,MeV and the neutron and proton cross section library for 10\,MeV$<$\textit{E}$<$200\,MeV when available. The standard flux parametrization used is from \cite{arm73}\cite{geh85}. Variation in the production rates are expected when different flux parametrizations or different  cross section libraries are used. Changes up to 200\% have been estimated. The expected number of events per tonne of material has been calculated for two cooling down scenarios after one year of activation at sea level. In order to reduce the cosmogenic-induced background after one year of exposure to a negligible level for a kilotonne scintillator based experiment it would be desirable to have an initial purification followed by cooling down time UG to remove the short lived isotopes.\\
The estimated in-situ underground activation due to muon induced neutrons for the long-lived, high Q-value isotopes at the SNOLAB location, 2 km underground, is less than 1 event/(yr$\cdot$t).

\section*{Acknowledgment}
The work was supported by the Deutsche Forschungsgemeinschaft (DFG). The work was performed using the resources of the center for information services and high performance computing (ZIH) at TU Dresden.\\
I thank J. Klein, S. Biller, K. Zuber, J.J. Back and  S. Cebri\'{a}n for valuable discussions. I acknowledge A. F. Barghouty for providing the YIELDX routine and the useful discussions. I thank E. V\'{a}zquez-J\'{a}uregui for the help with the rock neutron flux.\\ 
I specially thank N. Tolich for the suggestion to investigate a wide range of cosmogenic-induced backgrounds that is at the base of this paper. 

\end{document}